\documentstyle[12pt,epsf,blois]{article}

\begin{document}
\heading{ELEMENTAL ABUNDANCES AT EARLY TIMES: THE NATURE OF DAMPED LYMAN-ALPHA
SYSTEMS}

\author{Mercedes Moll\'{a}$^{1}$, Angeles I. D\'{\i}az$^{2}$, \& 
Federico Ferrini $^{3}$}
{$^{1}$
D\'epartement de Physique, Universit\'e Laval, Qu\'ebec, Qc, G1K~7P4, 
Canada.} {$^{2}$
Departamento de F\'{\i}sica Te\'{o}rica,
Universidad Aut\'{o}noma de Madrid, 28049 Madrid- Spain.} {$^{3}$ Dipartimento
 di Fisica, Universit\`{a} di Pisa, Piazza Torricelli 2, 56100 Pisa, Italia.}

\begin{bloisabstract}
The distribution of element abundances with redshift in 
Damped Ly$\alpha$ (DLA) systems can be adequately reproduced by the same 
model reproducing the halo and disk components of the Milky 
Way Galaxy at different galactocentric distances:  DLA 
systems are well represented by normal spiral galaxies in their early 
evolutionary stages.

\end{bloisabstract}

\section{Introduction}

Chemical evolution models have been developed to reproduce the
observational features obtained from the solar neighbourhood and
the whole Galaxy structure studies. These data give us information
about the galactic evolution up to a galactocentric distance of $\sim
15-25$ kpc and in the last $\sim 11-16$ Gyr.  However, in many cases,
models which give similar results for the present time diverge
significantly at different epochs.  New data on high
redshift objects -- {\em i.e.} objects far away in space and time --
can now be used to test model predicted past galaxy properties which
can then constitute important constraints for evolution models.

We use the multiphase model results calculated for MWG and
compare them with observational data from \cite{LSBCV}, and
\cite{PKSH}, in order to test if damped Ly$\alpha$ systems have
abundances consistent with those expected for high redshift spiral
galaxies.  We have assumed values of $q_0$ and $H_0$ of 0.1 and 50 km
s$^{-1}$ Mps$^{-1}$ respectively. The redshift $z$ corresponds to the
redshift of damped Ly$\alpha$ systems.

\section{The chemical evolution model}
\begin{figure*}
\epsfxsize=17cm
\epsfysize=17cm

\epsfbox{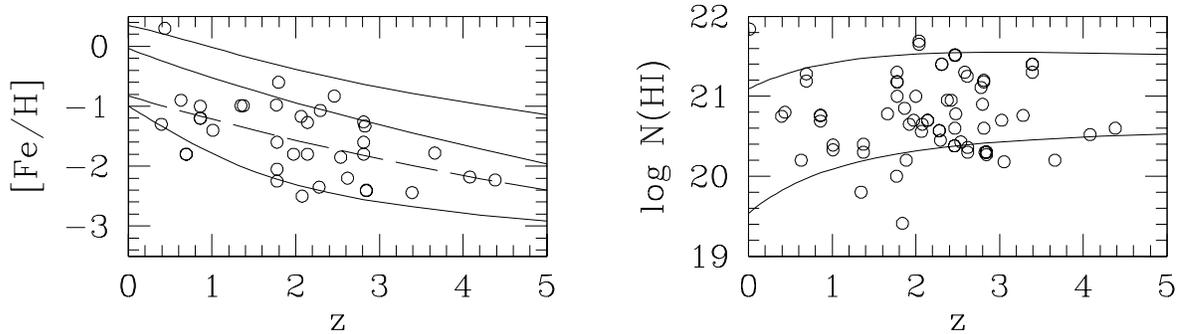}
\vspace*{-12cm}
\caption{ Left) [Fe/H] versus redshift. Solid lines are the  results
 for the three defined  regions. 
Right)  Column density for face--on and edge--on
 Milky Way twins. Circles are data from [2] and [4] }

\end{figure*}

We use the evolutionary history of the Milky Way, assumed to be a
typical spiral galaxy.  We take the multiphase model results from
\cite{FMPD} for MWG, where we computed the element distributions, in
the halo and in the disk, by varying the radial galactocentric distance
of every considered region.  We have extracted the outputs for the regions
centered on 8 kpc ({\em solar}) and 5 kpc and 14 kpc (here-on
called {\em inner} and {\em outer} zones) from the galactic center.
Assuming that the absorbing gas in the line of sight to the damped
Ly$\alpha$ systems may belong to both halo and disk components, we
evaluate the abundance by averaging the abundances for the two populations
with their corresponding weights.

Our results are shown in Figure 1. In the left panel, we show the average
abundance for the three defined regions.  The inner and  solar
region evolutions fit better the higher abundance data points, while the
outer region abundance fits the lower observed abundances.  The
history of the different regions can reproduce the spread in the
observations.  Observations possibly average also over a great part of
the Galactic structure: the dashed line shows our average for both
disk and halo at all radii.  In the right panel we plot the gas column density
results which also have contributions from the gas present
both in the spheroidal and disk components of the galaxy.  The column
density is sensitive to the inclination angle of the disk respect to
the line of sight: the two solid lines represent the evolution for a
face--on  galaxy (lower line) and for an edge--on  galaxy (upper line).  The distribution of the 
observed values is bound by these simple estimates.

\section{Conclusions}

The possibility that Damped Lyman $\alpha$ systems be indeed normal
spiral galaxies (\cite{W95}) must be considered: observations probably
refer to gas belonging to different components and therefore the
observed abundances are averaged over the halo and disk gas, if not
over the whole galactic content. We confirm that the
large metallicity spread observed in DLA systems can be explained by
the heterogeneous mixture of galactic regions and galaxy types.


%
\vfill
\end{document}